\documentclass{rspublic}

\usepackage{graphicx}
\usepackage{dcolumn}
\usepackage{bm}
\usepackage{epsfig}
\usepackage{amssymb}
\usepackage{psfrag}
\usepackage{natbib}
\usepackage{epsfig}
\usepackage{epstopdf}

\def\bea{\begin{eqnarray}}
\def\eea{\end{eqnarray}}
\def\beq{\begin{equation}}
\def\eeq{\end{equation}}

\begin{document}

\title{The nonlinear electromigration of analytes into confined spaces}[electromigration into confined spaces]
\author[Z. Chen and S.Ghosal]{ Zhen Chen and Sandip Ghosal}
\affiliation{Northwestern University, Department of Mechanical Engineering\\
2145 Sheridan Road, Evanston, IL 60208
}

\maketitle

\begin{abstract}{electromigration dispersion, electrokinetic injection, nonlinear waves}
We consider the problem of electromigration of a sample ion (analyte) within a uniform background electrolyte 
when the confining channel undergoes a sudden contraction. One example of such a situation arises in 
microfluidics in the electrokinetic injection of the analyte into a micro-capillary from a reservoir of much 
larger size.  Here the sample concentration propagates as a wave 
driven by the electric field. The dynamics is governed by the Nerst-Planck-Poisson system of equations for ionic transport.
A reduced one dimensional nonlinear equation describing the evolution of the sample concentration is derived.
We integrate this equation numerically to obtain the evolution of the wave shape and determine how the the injected 
mass depends on the sample concentration in the reservoir.
It is shown that due to the nonlinear coupling of the ionic concentrations and the electric field, the concentration 
of the injected sample could be substantially less than the concentration of the sample in the reservoir. 
\end{abstract}

\section{Introduction} 
\label{sec:Intro}
\begin{figure}[t]
  \includegraphics[width= 1.0\textwidth]{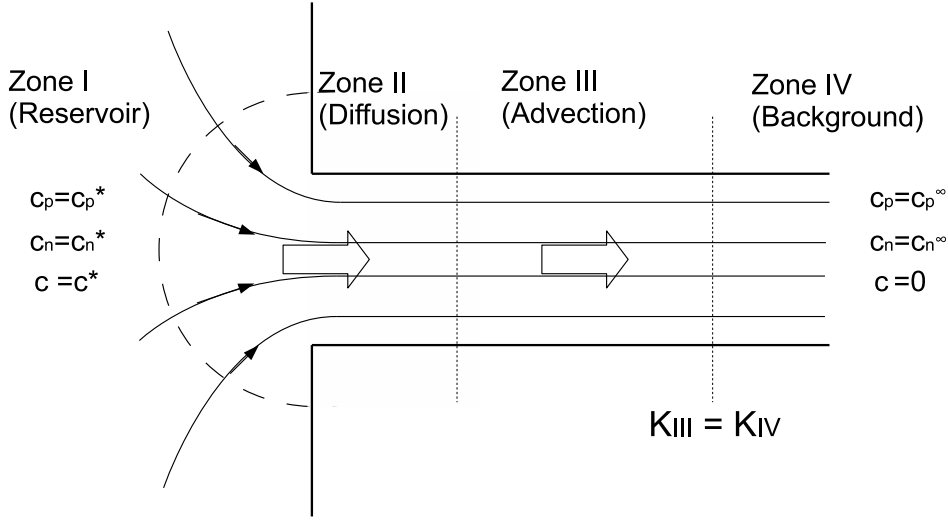}
\caption{Schematic diagram illustrating the electrokinetic injection of a sample 
(concentration $c^{*}$ in the reservoir) into a micro-channel}
\label{fig:Inject}       
\end{figure}

In microfluidic systems, electrokinetic injection is often employed  in order to insert the sample from a reservoir into a microfluidic channel~\citep{czebook1,czebook2}.
 A simplified problem depicting the electrokinetic injection process is shown in Figure~\ref{fig:Inject}. The reservoir initially contains the sample 
 dissolved in a background electrolyte of positive and negative ions. The channel contains only the background electrolyte. At time $t=0$,
 a voltage, $V$ is applied to an electrode in the reservoir located far from the inlet, while the channel outlet (assumed infinitely far away) 
 is electrically grounded. 
 We would like to describe how the concentration of the sample evolves with time
 as the sample moves into the micro-channel. For simplicity, we will assume that the capillary walls are uncharged so that there 
 is no electroosmotic flow.
 
 When the concentration of sample in the reservoir is low, the dynamics is linear and is described by an 
 advection diffusion equation with constant  advection speed.
 At high concentrations, the evolution is nonlinear due to the phenomenon of electromigration dispersion. 
The physical mechanism of electromigration dispersion may be 
explained~\citep{EMD}  in the following way: when the concentration of sample ions is significant 
in comparison to that of the background electrolyte, the electrical conductivity of the solution 
is locally altered. However, charge conservation and local electro neutrality requires the electric current
to be the same at all points along the axis of the capillary. Therefore, by Ohm's law, 
(ignoring, for the moment, the diffusive contributions to the current),
the electric field must change axially, since,  the product of the conductivity and electric field 
must remain constant. This varying electric field alters the effective migration speed of the sample ions, which in turn, alters 
its concentration distribution. The transport problem for sample ions then becomes nonlinear and shock like structures similar to those familiar from 
breaking water waves~\citep{whitam_book} can arise. 

The problem has a certain similarity 
with the classical ``dam break problem'' in hydrodynamics where a large stationary volume of water held back by a wall 
is suddenly released by  partial or complete removal of the restraining wall~\citep{whitham_effects_1955,stoker_48,ritter_1892}. However, in the present 
problem, the driving mechanism is the variation of the electric field along the channel instead of hydrostatic pressure variations
due to changes in wave height.

In the next section we present the mathematical formulation for a minimal problem consisting of three ionic species - the analyte, a co-ion and a counter-ion,
where the diffusivities of the three species are equal, the charge of the analyte is constant, and, the background electrolyte is fully dissociated. We will see that even in this highly idealized limit, the nonlinearities inherent in the system lead to surprising behavior. 
The idealizations permit reduction of the problem to the solution of a single partial differential equation in one dimension. In section~\ref{sec:numerics}, this equation is 
integrated numerically, and observations are presented for the behavior of the sample ion concentration as it moves into the micro-channel.
We find  that at high concentrations, the concentration of sample ions that enter the 
 micro-channel is significantly lower than that in the reservoir. In fact, as the concentration of sample ions in the reservoir is increased, the  concentration within the injected plug in the channel does 
not increase indefinitely; but, approaches a limiting value determined by the valence of the analyte ion and background electrolytes.
In section~\ref{sec:analysis}, we identify the reason for this behavior and use simple arguments based on conservation laws to deduce the 
limiting value of the sample concentration in the channel. This is compared to results from the numerical simulations and good 
agreement is found. Our findings are summarized in section~\ref{sec:conclusions} and its implications in a broader context as well as 
its limitations  are discussed.

\section{Problem Formulation}
 We consider a three ion system consisting of sample ions, co-ions and counter-ions all of equal diffusivity ($D$). 
It follows, from the Einstein relation, that the mobility ($u$) of the three ionic species are also the same. However,
the three species have different valences: $z_p,z_n,z$ and therefore different electrophoretic mobilities: $\mu_p = z_p e u,\mu_n = z_n e u,\mu = z e u$,
$e$ being the proton charge.
In this paper, the suffix $p$ will generally indicate the positive ion (cation), $n$ will indicate the negative ion (anion) and the absence of a 
suffix will indicate the sample ion (analyte). Thus, $z_p$ is positive, $z_n$ is negative and $z$ could be of either sign. 
The discussion in the rest of this section follows closely the earlier work of \cite{EMD}. 

We consider the simplest situation where the ions migrate from a semi-infinite reservoir to a uniform channel 
with no wall charge. Then the problem is entirely one dimensional, and, the coupled equations describing the concentrations of the three ion species are
\begin{eqnarray} 
\frac{\partial c_{p}}{\partial t} + \frac{\partial}{\partial x} \left( z_{p} euE c_{p} \right) &=& D \frac{\partial^{2} c_{p}}{\partial x^{2}}, \label{T+}\\
\frac{\partial c_{n}}{\partial t} + \frac{\partial}{\partial x} \left( z_{n} euE c_{n} \right) &=& D \frac{\partial^{2} c_{n}}{\partial x^{2}}, \label{T-}\\
\frac{\partial c}{\partial t} + \frac{\partial}{\partial x} \left( zeuE c \right) &=& D \frac{\partial^{2} c}{\partial x^{2}}, \label{T}
\end{eqnarray}
where $E$ is the local electric field, $x$ is the distance along the capillary and $t$ is time. We will assume that 
the x-axis points in the direction of front motion from the reservoir into the capillary.
  Since characteristic spatial scales are always much larger than 
the Debye length, local electro-neutrality holds~\citep{PRSA10}. Thus,
\begin{equation} 
z_{p}c_{p}+z_{n} c_{n}+zc = 0.
\label{LEN}
\end{equation}

If we multiply equations (\ref{T+})-(\ref{T}) by the respective ionic charges, sum them, and use equation (\ref{LEN}), we get an 
equation that describes the constancy of electric current 
\begin{equation} 
 \frac{\partial}{\partial x} \left[ e^{2}u ( z_{p}^{2} c_{p} + z_{n}^{2} c_{n} + z^{2} c ) E \right] = 0.
 \label{conserve_curr}
 \end{equation} 
Note, that, the net contribution from the diffusive fluxes vanish exactly on account of the assumption of equal diffusivity of ions 
and local electro-neutrality.
Equation (\ref{conserve_curr}) may then be integrated to yield
\begin{equation} 
 ( z_{p}^{2} c_{p} + z_{n}^{2} c_{n} + z^{2} c ) E = ( z_{p}^{2} c_{p}^{\infty} + z_{n}^{2} c_{n}^{\infty}) E_{\infty}, 
  \label{current}
 \end{equation}
 where $E_{\infty}$ is the electric field in the capillary very far away from the inlet, and, $c_{n}^{\infty},c_{p}^{\infty}$ are respectively 
 the corresponding negative and positive ion concentrations in the background electrolyte. 
 Clearly, $c_{p}^{\infty}$ and $c_{n}^{\infty}$ are not independent but are related by the electro-neutrality condition 
 \begin{equation}
 z_{p}c_{p}^{\infty} +z_{n} c_{n}^{\infty} = 0. \label{LEN_bge}
 \end{equation}
 If we now introduce the Kohlrausch function~\citep{kohlrausch}
 \begin{equation} 
 K(x,t) = (c_{p}+c_{n}+c)/u, \label{defineK}
 \end{equation}
 then it follows from equations~(\ref{T+})-(\ref{T}) and (\ref{LEN}) that $K$ satisfies the diffusion equation 
 \begin{equation} 
 \frac{\partial K}{\partial t} = D \frac{\partial^{2} K}{\partial x^{2}}. \label{dK/dt}
 \end{equation} 
 The three algebraic relations: (\ref{LEN}), (\ref{current}) and (\ref{defineK}), may be used to 
 express all four dependent variables in the 
problem: $c_{p},c_{n},c$ and $E$ in terms of any one of them and the function $K(x,t)$.
We choose to express all the variables in terms of $c$ and $K$. These relations take a particularly 
simple form if we use the normalized concentration 
\begin{equation} 
\phi = \frac{c}{c_{n}^{\infty} }
\end{equation} 
and 
\begin{equation} 
\theta (x,t) = \frac{u z_{p}}{z_{p}-z_{n}} \frac{K(x,t)}{c_{n}^{\infty}}
\end{equation}
In terms of these variables, we then have
\begin{eqnarray} 
E &=& \frac{E_{\infty}}{\theta - \alpha \phi}, \label{eq4E}\\
\phi_{p} &=& \frac{ c_{p} }{ c_{n}^{ \infty } } = - \frac{z-z_n}{z_p - z_n} \phi - \frac{z_n}{z_p}  \theta, \label{eq4phip} \\
\phi_{n} &=& \frac{ c_{n} }{ c_{n}^{\infty} } = \frac{z-z_p}{z_p - z_n} \phi +  \theta, \label{eq4phin}
\end{eqnarray}
so that, equation (\ref{T}) gives the following evolution equation for $\phi$:
\begin{equation}
\frac{\partial \phi}{\partial t} + \frac{\partial}{\partial x} \left( \frac{v \phi}{\theta - \alpha \phi} \right) = D \frac{\partial^{2} \phi}{\partial x^{2}}  \label{pde_4_phi},
\end{equation} 
where $v= ze u E_{\infty}$ is the velocity of an isolated sample ion in the constant  electric field $E=E_{\infty}$, and, 
following the notation of \cite{EMD}, $\alpha = [(z-z_n)(z-z_p)/z_n(z_p-z_n)]$, is the ``velocity-slope parameter''.

Since $\theta$ is proportional to $K$, it is also a passive scalar. The initial and boundary conditions on $\theta$
in the domain $ 0 \leq x < \infty$  are
\begin{eqnarray}
\theta(x,0) &=& 1 \quad \mbox{ if $x>0$ },\\
\theta(0,t) &=& \theta^{*},\\
\theta (\infty,t) &=& 1,
\end{eqnarray}
where $ \theta^{*}$ is the value of $\theta$ in the reservoir. The reservoir is considered to be ``well mixed'' so that all ionic 
concentrations remain constant within it. An exact solution for $\theta$ may then be found using Fourier's method,
\begin{equation}
\theta = \theta^{*} + (1 - \theta^{*} ) \mbox{erf} \left( \frac{x}{\sqrt{4Dt}} \right),  \label{soln_theta}
\end{equation}
where 
\begin{equation} 
\mbox{erf}(x) = \frac{2}{\sqrt{\pi}} \int_{0}^{x} e^{-y^{2}} \; dy
\end{equation} 
is the error function. 
Since the concentration of sample in the reservoir is being held constant  at some value $\phi = \phi^{*}$, 
we have 
\begin{equation} 
\theta^{*} = \frac{c_{n}^{*}}{c_{n}^{\infty}} + \frac{z_{p} - z}{z_{p}-z_{n}} \phi^{*}.
\end{equation} 
The superscript $*$ indicates the value of the corresponding variable in the reservoir. Thus, $c_{n}^{*}$ is the concentration of negative ions in the reservoir. 

Equation~(\ref{pde_4_phi}), on the other hand, does not readily admit an analytical solution, since, unlike the equation for $\theta$,
it cannot be reduced to an ordinary differential equation by means of the similarity variable $\eta = x / \sqrt{4 Dt}$. 
We will therefore obtain the time evolution of the normalized sample concentration $\phi(x,t)$ by substituting equation~(\ref{soln_theta}) in equation~(\ref{pde_4_phi}) 
and integrating the resulting one dimensional partial differential equation numerically. Before we present this solution in the next section, it is useful to note
that at large distances from the reservoir, or more precisely, if $x \gg \sqrt{4 Dt}$, equation~(\ref{soln_theta}) implies that $\theta \sim 1$, so that 
equation~(\ref{pde_4_phi}) reduces to 
\begin{equation}
\frac{\partial \phi}{\partial t} + \frac{\partial}{\partial x} \left( \frac{v \phi}{1 - \alpha \phi} \right) = D \frac{\partial^{2} \phi}{\partial x^{2}}  \label{pde1_4_phi}
\end{equation} 
which was studied earlier by \cite{EMD}.

\section{Numerical Simulations}
\label{sec:numerics}
We will find it convenient to express our results 
in terms of a characteristic  length, $w$ (the channel width), which gives a characteristic time $w/v$.
Then clearly, the only parameters in the problem are ${\mbox Pe} = vw/D$, which may be regarded 
as a ``P\'{e}clet number'' based on the electromigration speed, $v$, or a ``field strength parameter''; the two valence ratios 
$z_n/z,z_p/z$; and, the parameter $\theta^{*}$ characterizing the ionic composition of the mixture in the reservoir.
For definiteness, we will assume that the sample is a cation ($z>0$), and that 
\begin{equation} 
c_{p}^{*} = c_{p}^{\infty}
\end{equation}
This will be true, for example, if the electrolyte in the reservoir is prepared by adding the sample in solid form (so there is no dilution)
to a stock solution of the background electrolyte that fills the  micro-channel (since the sample contributes only counter-ions, the co-ion concentration is not altered).
Using the electro-neutrality conditions, 
\begin{equation}
\theta^{*} = 1 - \frac{z_p (z - z_n)}{z_n (z_p - z_n)} \phi^{*}.
\end{equation}
The initial and boundary conditions for $\phi$ are
\begin{eqnarray}
\phi(x,0) &=& 0, \quad \mbox{if $ x >0$}\\
\phi(0,t) &=& \phi^{*},\\
\phi (\infty,t) &=& 0.
\end{eqnarray}
We keep the valence ratios fixed at $z_p/z=2$, $z_n/z=-1$.

\begin{figure}[t]
  \includegraphics[width= 1.0\textwidth]{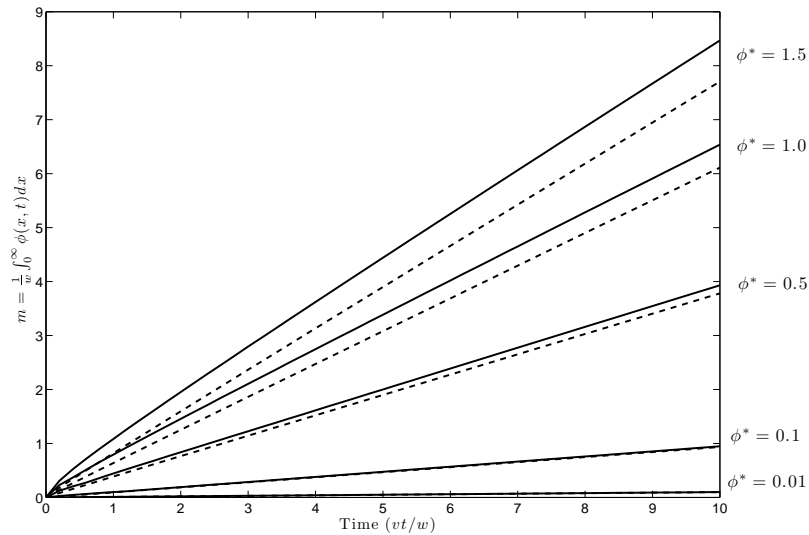} 
 \caption{The increase of sample ions in the micro-channnel as a function of dimensionless 
 time at different dimensionless reservoir concentrations, $\phi^{*}$. The solid line corresponds to $\mbox{Pe}= 10$ and the dashed 
 line corresponds to $\mbox{Pe}=200$. }
\label{fig:m}       
\end{figure}
Since we have reduced our problem to a one  dimensional one, the numerical integration is quite straightforward.
We used a  finite volume method to discretize equation (\ref{pde_4_phi})  in space using 
an adaptive grid refinement algorithm that is enabled by applying the Matlab library ``MatMOL''~\citep{vande_wouwer_simulation_2004}.
The spatially discretized system of equations is then integrated in time using the Matlab solver ``ode45''~\citep{shampine_matlab_1997}
which is based on an explicit Runge-Kutta (4,5) formula. 

\begin{figure}[t]
  \includegraphics[width= 1.0\textwidth]{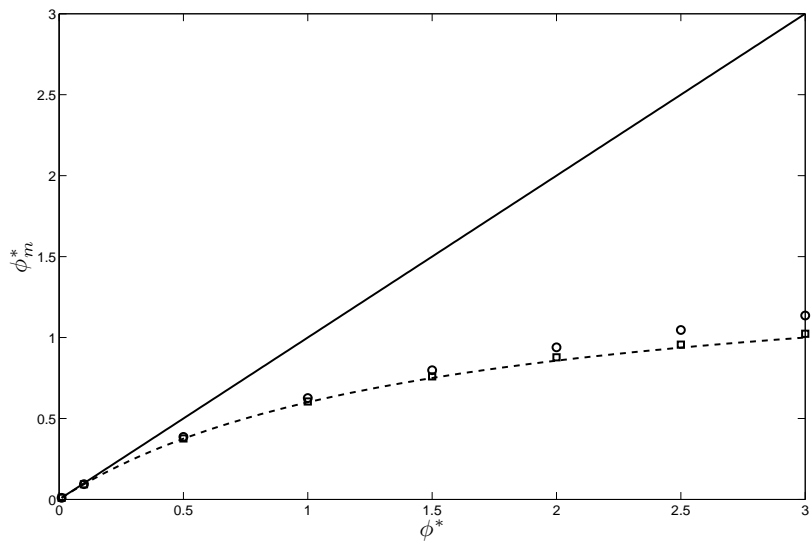}
\caption{The normalized concentration of analyte in the micro-channel ($\phi_{m}^{*}$) as a function of the 
corresponding quantity in the reservoir ($\phi^{*}$). The solid line indicates the result expected from linear theory, the dashed line is the result 
predicted by the nonlinear theory, equation~(\ref{eq:phi_inter}). The symbols (circle: Pe=10,  square: Pe=200) are obtained by numerical integration of equation~(\ref{pde_4_phi}). }
\label{fig:m_rate}       
\end{figure}
Figure~\ref{fig:m} shows the amount of sample that has moved into the reservoir at time $t$, defined as 
\begin{equation} 
m(t) = \frac{1}{w} \int_{0}^{\infty} \phi(x,t) \; dx.
\end{equation}
Simulations are shown for ${\mbox Pe}=10$ and $200$. At long times, or more precisely, 
if $vt/w \gg {\mbox Pe}^{-1}$, the expected asymptotic form for $m(t)$ is 
\begin{equation} 
m(t) \sim ( v \phi^{*} / w ) t. \label{linear_flux}
\end{equation}
Figure~\ref{fig:m} shows that indeed, at large times, $m(t) \propto t$. We determined the slope of the $m(t)$ curve at large times and plotted  
the dimensionless quantity $w \dot{m} / v$ as a function of $\phi^{*}$ in Figure~\ref{fig:m_rate}. It is seen that $w \dot{m} / v \sim \phi^{*}$ 
as long as $\phi^{*} \ll 1$ but as the concentration of sample in the reservoir is increased, the injection rate does not increase 
indefinitely but tends to saturate to a limiting value, that is, $w \dot{m} / v \sim \phi_{m}^{*} < \phi^{*}$.

\begin{figure}[t]
  \includegraphics[width= 1.0\textwidth]{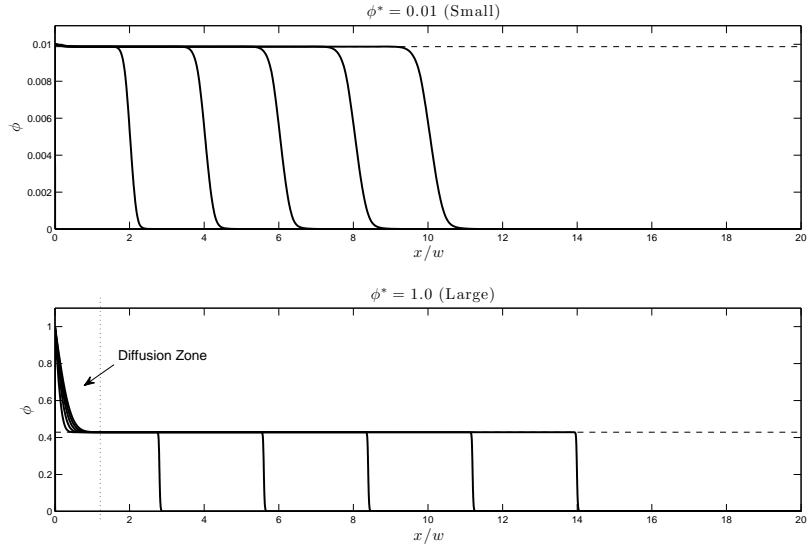}
\caption{Concentration profiles in the micro-capillary at successive times: $vt/w=2,4,6,8,10$. If the reservoir concentration is large 
(lower panel), a concentration jump develops at the entrance of the capillary, with a narrow diffusion zone connecting the values in the 
reservoir with the much smaller value of the concentration within the capillary. The concentration jump is insignificant if the reservoir concentration is small (upper panel). The dashed line is $\phi_{m}^{*}$ evaluated using equation~(\ref{phi_m_th}). Here Pe=200.}
\label{fig:c_compare}       
\end{figure}

The reason for this saturation of the injection rate may be understood by examining the concentration profiles $\phi(x,t)$ 
shown in Figure~\ref{fig:c_compare}. It is seen, that as the sample moves into the capillary, the concentration in the capillary is approximately equal to the 
the concentration in the reservoir as long as this concentration remains small (upper panel, $\phi^{*} = 0.01$). However, if the 
concentration in the reservoir is large (lower panel, $\phi^{*} = 1.0$) the concentration in the channel approaches a limiting value 
$\phi^{*}_{m}$ independent of the reservoir concentration. 

\begin{figure}[t]
  \includegraphics[width= 1.0\textwidth]{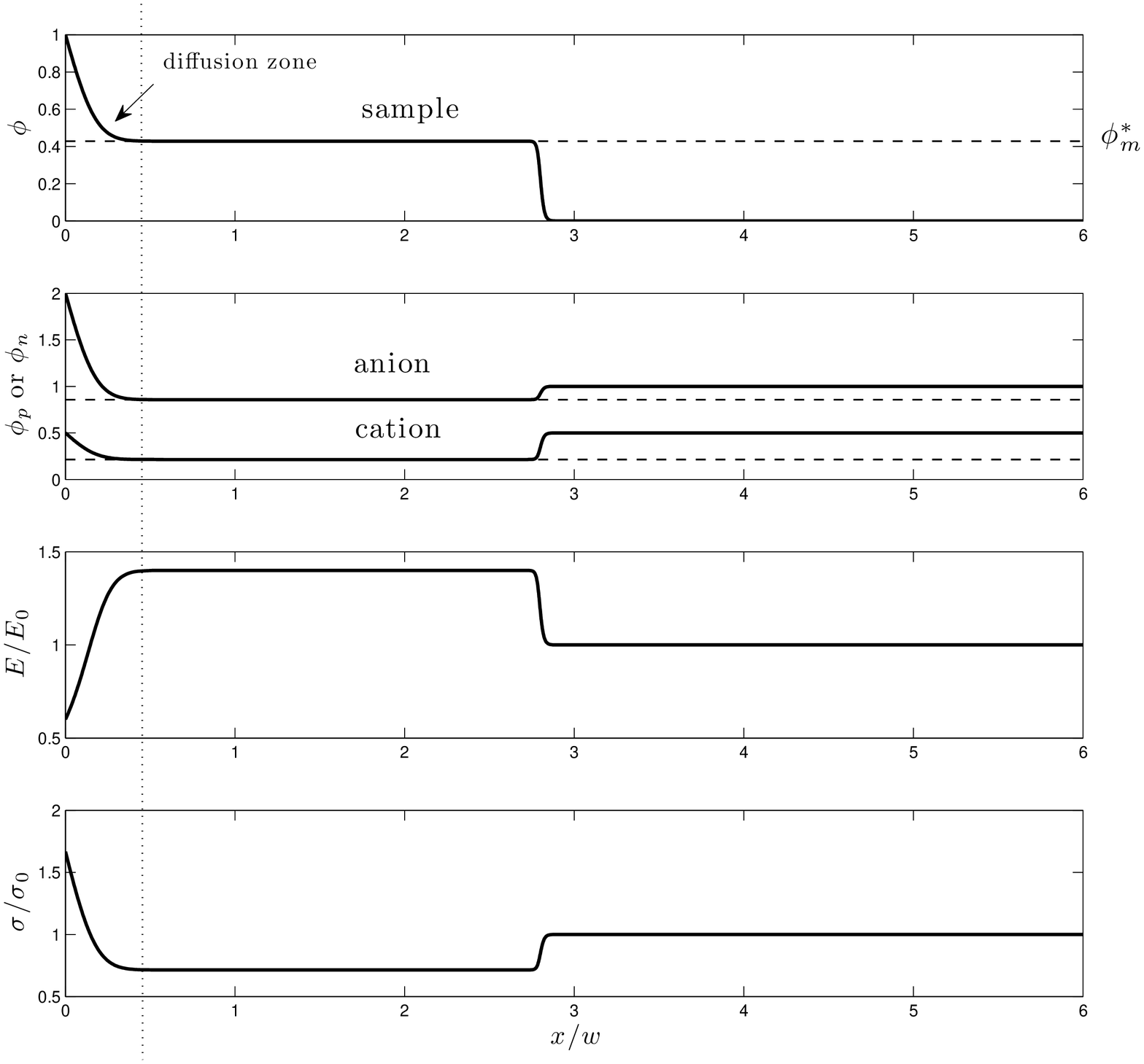}
\caption{Same as in Figure~\ref{fig:c_compare} except that profiles of all dependent variables are 
shown but at a fixed time instant, $vt/w=2$. The dashed lines are the concentrations in the capillary 
calculated using equation~(\ref{eq:phi_inter}). The bottom panel shows the electric conductivity ($\sigma$) normalized 
by its downstream value ($\sigma_{0}$).}
\label{fig:c_E}       
\end{figure}
In Figure~\ref{fig:c_E}, we have reproduced the concentration distribution $\phi(x,t)$ that appears in the lower panel of 
Figure~\ref{fig:c_compare}, 
together with the concentrations of the other ionic species $\phi_p (x,t)$ and $\phi_n(x,t)$, the local electric field $E(x,t)$,
and, the local electrical conductivity due to ions, $\sigma (x,t)$.
These quantities are obtained from equations (\ref{eq4E}), (\ref{eq4phip}) and (\ref{eq4phin}).  The mechanism for the 
reduction of the concentration from $\phi = \phi^{*}$ in the reservoir to $\phi = \phi^{*}_{m} < \phi^{*}$ in the capillary 
may be understood from these graphs. At large values of the sample concentrations in the reservoir,  there is a a sharp rise 
in the solution conductivity in the immediate vicinity of the inlet resulting in a drop in the electric field. The flux of sample into the 
micro-channel is then determined by a combination of this greatly reduced electromigrational flux and the diffusive flux and this is less 
than the expected flux indicated by equation (\ref{linear_flux}), which should hold in the linear regime.

\section{Analysis} 
\label{sec:analysis}

An approximate theoretical determination of the critical concentration  $\phi^{*}_{m}$ may be provided using the conservation 
equations. The method of doing this is in fact entirely analogous to that of the ``Moving Boundary Equations'' (MBE)~\citep{dole_theory_1945} 
for describing advancing fronts (e.g. in isotachophoresis). The conceptual framework is 
illustrated in Figure~\ref{fig:Inject}. The domain is decomposed into four parts: (I) the ``Reservoir Zone'' within the well mixed body of the 
reservoir where all concentrations are constant (II) the ``Diffusion Zone'' which is the initial 
part of the micro-channel of length of order $\sqrt{ 2 D t}$ -- here the ionic diffusive flux and the electromigrational  flux driven by the electric field 
are comparable, (III) the ``Advection Zone'' where diffusive fluxes are negligible (because concentration gradients are small) 
and ionic fluxes are primarily due to electromigration, and, (IV) the ``Background Zone'' far downstream where
the sample ions have not yet penetrated.

The boundary between the Zones (I) and (II) is stationary (and located at the channel inlet) 
whereas the boundaries between the other zones propagate to the right. 
The arrows in Figure~\ref{fig:Inject} indicate fluxes of ions across the stationary zone boundary. 
Equality of the fluxes for each species across the zone boundary require:
\begin{eqnarray}
	\pi r  \phi^{*} E^{(I)}(r)  &=& w E^{(III)}\phi^{*}_{m} \\
	\pi r \phi^{*}_n E^{(I)} (r)  &=& w E^{(III)}\phi^{(III)}_n 
\end{eqnarray}
where $E^{(I)}(r)$ represents the electric field in the reservoir at a radial distance $r$ from the inlet. 
The superscript (I), (II), (III) or (IV) indicate the zone in which the corresponding variable is evaluated.
Clearly $\phi^{(I)} = \phi^{*}$, $\phi_{n}^{(I)} = \phi_{n}^{*}$ and $\phi^{(III)} = \phi^{*}_{m}$.
Therefore, taking the ratio,
\begin{equation}
	\phi^{*}_{n} = ( \phi^{(III)}_n / \phi^{*}_{m} ) \phi^{*}
	\label{eq:phi_n}
\end{equation}
The Kohlrausch function, or, equivalently, $\theta(x,t)$, propagates only by diffusion. Therefore,
$\theta (x,t)$ differs from unity only in Zones (I) and (II). Thus, 
\begin{equation}
	\phi_{n}^{(III)} + \phi_{p}^{(III)} +  \phi_{m}^{*}  = 1 
	 \label{eq:Kinter}
\end{equation}
The electro-neutrality condition (valid in all zones) for Zone (III) is:
\begin{equation} 
z_p \phi_{p}^{(III)} + z_n \phi_n^{(III)} + z \phi_{m}^{*} = 0. 
\label{eq:Z3LEN}
\end{equation} 
By combining equations (\ref{eq:phi_n}) and (\ref{eq:Kinter}) and using the electro-neutrality condition, equation~(\ref{eq:Z3LEN}),
we get  an equation for determining $\phi_{m}^{*}$
\begin{equation}
	(1 - z/z_p) \phi_{m}^{*} + r (1 - z_n/z_p) \phi_{m}^{*}  = 1 - z_n/z_p,
\end{equation}
where the ratio $\phi^{*}_n / \phi^{*} = r$ is a constant determined by the ionic composition of the electrolyte in the reservoir. 
Solving the above linear equation for $\phi_{m}^{*}$ we have 
\begin{equation}
	\phi_{m}^{*} = \left[ r + (z_p - z)/(z_p - z_n) \right]^{-1}.
	\label{eq:phi_inter}
\end{equation}

In our numerical experiment we assume that the sample is a cation, and that the co-ion 
concentration in the reservoir is the same as in the channel, so that 
$\phi^{*}_p = \phi_{p}^{\infty} = -z_n/z_p$. The electro-neutrality condition then implies that  $\phi^{*}_n = 1 - (z/z_n) \phi^{*}$, so that 
equation (\ref{eq:phi_inter}) becomes
\begin{equation}
\phi_{m}^{*} = \frac{\phi^{*}}{1 + \xi \phi^{*}}  \label{phi_m_th}
\end{equation} 
where 
\begin{equation}
\xi = - \frac{z_p (z-z_n)}{z_n (z_p - z_n)}
\end{equation} 
is a positive number. An expression for $\theta^{*}$ in terms of the sample ion concentration, $\phi^{*}$, may also be derived by simple algebra:
\begin{equation} 
\theta^{*} = 1 - \frac{z_p (z-z_n)}{z_n (z_p - z_n)} \phi^{*}.
\end{equation} 
In Figure~\ref{fig:m_rate}, we used equation (\ref{phi_m_th}) to plot  $\phi_{m}^{*}$ as a function of 
 $\phi^{*}$. This curve is seen to describe very well 
the nonlinear saturation of the injected mass with increasing sample concentration in the reservoir.

The denominator in equation~(\ref{pde_4_phi}) vanishes if $\phi = \theta / \alpha$. Far from the 
inlet, the singularity is approached as $\phi \rightarrow 1/\alpha$ (if $\alpha > 0$), since, $\theta \sim 1$. 
It was shown earlier~\citep{EMD} that the requirement of positivity of all ionic 
concentrations imposes a condition of self-consistency, 
 $\phi(x,t)< \phi_{c}$  where
\begin{eqnarray} 
\phi_{c} = \left\{
\begin{array}{ll}
(z_p - z_n)/(z_p -z)  & \mbox{if $z>0$}\\
 -  [ z_n (z_p-z_n) ] / [ z_p (z-z_n) ] & \mbox{if $z<0$}.
 \end{array}
 \right.
\end{eqnarray}
Since, $\phi_{c} < 1/\alpha$, if $\alpha >0$ (see Appendix), the singularity is never reached 
if initially $\phi$ is less than $\phi_{c}$ everywhere in the channel. 
This, however, leaves open the possibility that the initial condition $\phi(0,t)$ might
be such, that, $\phi$ exceeds $\phi_{c}$ in some parts of the domain. 
We have now shown that if electrokinetic injection is used to insert the sample 
into the capillary, the shock at the inlet 
ensures that the injected sample concentration does not exceed $\phi_{m}^{*} < \phi_{c}$  (see Appendix), no matter 
how high the sample concentration in the reservoir may be. In capillary zone electrophoresis, one chooses the peak shape after 
electrokinetic sample injection as the ``initial condition''. Then, the only initial conditions possible  are the 
``realizable'' kind, where,  $\phi(x,0) < \phi_{c}$. 

The speed of propagation of the analyte front in Figure~\ref{fig:c_compare} may be readily calculated when the front is far from the inlet.
To do this, we note that since ${\mbox Pe} = 200 \gg 1$ and $\theta \rightarrow 1$, equation~(\ref{pde1_4_phi}) may be written as 
\begin{equation}
\frac{\partial \phi}{\partial t} + \frac{\partial Q}{\partial x}  = 0  \label{pde1a_4_phi}
\end{equation} 
where 
\begin{equation}
Q(\phi) = \frac{v \phi}{1 - \alpha \phi}.
\end{equation} 
Then, the shock propagation speed is given by~\citep{whitam_book} the Rankine-Hugoniot jump condition 
\begin{equation} 
V_{s} = \frac{Q(\phi_{2})-Q(\phi_{1})}{\phi_{2} - \phi_{1}}
\end{equation} 
where the subscripts $2$ and $1$ refer to conditions just behind the shock and just ahead of the shock 
respectively. Putting $\phi_{1} = 0$ and $\phi_{2} = \phi_{m}^{*}$, we find, for the conditions shown in the 
lower panel of Figure~\ref{fig:c_compare}, $\phi_{m}^{*} = 3/7 \approx 0.43$, so that, $V_{s} = 1.405 \: v$.
The shock speed determined from direct measurement of the front displacement in Figure~\ref{fig:c_compare} 
is $1.4 \: v$, in close agreement with the theoretical prediction. In the upper panel, $\phi^{*} = 0.01$, so that $\phi_{m}^{*} \approx \phi^{*}$.
In this case, both the theoretical and the measured values give $V_{s} \approx v$.

\section{Conclusions}
\label{sec:conclusions}

The problem of electromigration of a sample ion from a large reservoir into a confined channel 
in the presence of a fully dissociated background electrolyte was studied in the idealized 
situation where all ionic species have the 
same diffusivity.  If the sample ion concentration is small compared to that of the background electrolyte,
the behavior of the system is exactly as one might expect. When the electric field is switched on, a sample 
plug with concentration equal to that in the reservoir, is drawn into the microchannel. However, if the 
sample ion concentration is comparable or large  compared to the background electrolyte, the behavior 
of the system is dominated by the intrinsic nonlinearity of the electro-diffusional problem, and the 
outcome is quite counter-intuitive. Once again, the sample is drawn into the microchannel as a plug 
of increasing length, but the concentration of sample ions in the plug is less than in the reservoir. 
In fact, as the reservoir concentration is progressively increased, the concentration of sample ions 
in the capillary approaches a limit that depends on the valences of the three ionic species but is independent 
of the reservoir concentration. 

In an earlier paper~\citep{EMD}, the authors presented a theory of electromigration of a sample in an infinitely long  micro-channel 
when only three ions of equal diffusivity are present. There, it was shown, that, the requirement that 
none of the three ionic concentrations could ever be negative, put a limit to the validity of the theory. 
It was therefore required that the initial condition be such, that, the dimensionless 
sample concentration not exceed a certain (positive) critical concentration, $\phi_{c}$. 
Here we have shown that initial conditions that do not satisfy this requirement cannot arise 
 if the analyte is introduced into the 
micro-capillary by elecrokinetic injection. In fact, the diffusional boundary layer at the channel entrance 
ensures that the concentration of analyte drawn into the micro-capillary is always less than $\phi_{c}$.
Thus, the critical concentration can never be exceeed if the sample is introduced by electrokinetic injection,
and consequently, the singularity at $\phi = 1/\alpha$ (if $\alpha >0$), inherent in equation~(\ref{pde1_4_phi}) 
is never reached. 

One may question whether the strongly nonlinear regime considered here is of relevance to actual laboratory practice. The answer
depends on the numerical value of the critical concentrations, $\phi_{m}^{*}$ and $\phi_{c}$.
If the sample and carrier ions have similar valences, then these critical concentrations are all of order unity. Thus, to 
approach these critical values, the sample ions in the injected plug will need to be present at concentrations approaching 
that of the carrier electrolyte. Such high concentrations are normally not employed in laboratory practice with capillary 
electrophoresis. However, if the sample 
is a macro-ion the critical values may actually be quite small. For example, at pH 2.0 Bovine serum albumin has a 
valence, $z \sim 55$~\citep{ford_measurement_1982}. Then, in a univalent 
carrier electrolyte we have $\phi_{c} \sim 0.04$, so that the strongly nonlinear regime studied here may be easily reached.

The shock like transition in concentration at the entrance of the micro-channel observed here is reminiscent of ``de-salination shocks''
that propagate outward from micro-channel/nano-channel junctions~\citep{mani_zangle_santiago_2009,zangle_mani_santiago_2009}
or from the surface of a perm-selective membrane embedded in a microfluidic channel. The mechanism of these shocks is related to the 
nonlinear coupling between the ion depletion zone (due to concentration polarization) near the membrane with the electric Debye layers at the walls 
of the micro-channel. The problem considered here is much simpler; Debye layers or concentration polarization are not involved. Nevertheless,
it is another example of shock-like behavior that may be traced to the nonlinear nature of the underlying Nernst-Planck-Poisson 
system of equations for ionic transport. As in the case of de-salination shocks, the effect described here would also be relevant 
for solutes electromigrating into porous media~\citep{mani_bazant_2011} or electromigration of analytes through variable 
geometry channels, such as pores in membranes, where lubrication theory~\citep{gh_02c} can be used to generalize the analysis 
presented here. Finally, if the channel walls are charged, electroosmotic flow would arise so that the transport problem 
becomes intrinsically two or three dimensional. In such cases, homogenization can be used to achieve a reduced description, as 
shown by \cite{gh_ch_12} for electromigration in an infinitely long uniform channel.\\[2ex]

\noindent {\em Acknowledgement:}  Support from the National Institute of Health (NIH) under grant R01EB007596 is gratefully acknowledged.

%
%




\renewcommand{\theequation}{A-\arabic{equation}}
  \setcounter{equation}{0}  
 \section*{Appendix: Proof of the inequality $\phi^{*}_{m} < \phi_{c} < 1/\alpha$}

If $\alpha >0$, $z_p > z > z_n$. Then 
\begin{equation}
	\alpha \phi_c  = \left\{
	\begin{array}{ll}
	\frac{z_p - z_n}{z_p - z}  \frac{(z-z_n)(z - z_p)}{z_n (z_p - z_n )} = \frac{-z_n + z}{-z_n} < 1& \mbox{if $z<0$}\\
	- \frac{z_n}{z_p} \cdot \frac{z_p-z_n}{z - z_n} \frac{(z-z_n)(z - z_p)}{z_n (z_p - z_n )} = \frac{z_p - z}{z_p} < 1& \mbox{if $z>0$}
	\end{array}
		\right.
\end{equation}
thus, $\phi_{c} < 1/\alpha$ if $\alpha >0$. 

To prove the first part of the  inequality, $\phi^{*}_{m}  < \phi_{c}$, we first show that $r > -z /z_n$. 
This is clearly true if $z<0$. To show this when $z>0$,
first  we use the electro-neutrality condition to express $\phi_{p}^{*}$ in terms of the other variables:
\begin{equation} 
\phi_{p}^{*} = - \frac{z}{z_p} \phi^{*} - \frac{z_n}{z_p} \phi_{n}^{*}  = \frac{\phi^{*}}{z_p} (- z - r z_n ).
\end{equation} 
Now we must have $\phi_{p}^{*} > 0$. This is always true if $z<0$, but if $z>0$, we require that $r > - z/z_n$. 

We will now show that $\phi^{*}_{m}  < \phi_{c}$ by considering the two cases $z>0$ and $z<0$ separately.
First suppose that $z <0$. Then 
\begin{equation} 
\phi^{*}_{m} = \frac{1}{ r + (z_p - z)/(z_p - z_n) } < \frac{1}{ (z_p - z)/(z_p - z_n) } = \frac{z_p - z_n}{z_p - z} = \phi_{c}
\end{equation}
Now suppose that $z >0$. Then 
\begin{eqnarray} 
\phi^{*}_{m} &=& \frac{1}{ r + (z_p - z)/(z_p - z_n) } \\
&<&  \frac{1}{ -(z/z_n) + (z_p - z)/(z_p - z_n) } = - \frac{z_n}{z_p} \frac{z_p - z_n}{z - z_n} = \phi_{c}
\end{eqnarray}
Thus, in all cases, $\phi^{*}_{m} < \phi_{c}$ which completes the proof.
\end{document}